\documentclass[twocolumn,showpacs,preprintnumbers,superscriptaddress,amsmath,amssymb]{revtex4}
\usepackage{amsfonts}
\usepackage{amsmath}
\usepackage{bbm}
\usepackage{amssymb}
\usepackage{graphicx}%
\usepackage{footmisc}
\usepackage{bm}
\usepackage{braket}
\usepackage{CJK}
\usepackage{hyperref}
\usepackage{cleveref}
\crefname{equation}{Eq.}{Eqs.}

\newcommand{\abs}[1]{\lvert#1\rvert}
\newcommand{\norm}[1]{\lVert#1\rVert}
\newcommand{\Tr}{\text{Tr}}

\newcommand{\Max}{\text{Max}}
\newcommand{\Eq}[1]{Eq. (#1)}
\begin{document}
\begin{CJK*}{UTF8}{gbsn}
\title{N-Z equation from assignment maps }

\author{Zhiqiang Huang (黄志强)}
\email{hzq@wipm.ac.cn}
\affiliation{Wuhan Institute of Physics and Mathematics, Chinese Academy of Sciences, Wuhan 430071, China}

\date{\today}

\begin{abstract}
The Nakajima-Zwanzig (N-Z) equation is a powerful tool to analyze open quantum systems and  non-Markovian behavior. In this paper, we rewrite the N-Z equation with  assignment maps. This approach extends the application to situations with bipartite state systems and environments with nonnegligible discord initially and/or during the dynamics. We apply the new equation in the quantum lattice model. The bound of the influence of the inhomogeneous terms reflects how the scale of the environment impacts the non-Markovian behavior. Under a large-scale environment,  with some conditions, the new equation is applicable even if the system and environment are maximally entangled  initially and/or during the dynamics. 
\end{abstract}


\maketitle
\end{CJK*}

\section{INTRODUCTION}
In both experimental and theoretical research, people are often interested in only specific  parts of objects, but there are always interactions between the parts of the whole.  Hence, the open system is common and important. A detailed review of open quantum systems can be found in\cite{BP02,VA17}. In many cases, the dynamics satisfy the Markov approximation, which makes such processes history independent. In quantum dynamics, the evolution of such systems is described by the well-known Lindblad equation. As our ability to observe and control quantum systems increases, non-Markovian behavior becomes increasingly important. The N-Z equation is a powerful tool to analyze open quantum systems.  With the Markov approximation, the N-Z equation can provide a Lindblad master equation.
 Without this approximation, the equation can describe non-Markovian behavior. Thus, the N-Z equation is widely used in the research of open systems. 

The N-Z equation uses the projection-operator method to derive the dynamic equation of the system.  The original projection-operator method maps the total density matrix to a tensor product state, and the irrelevant part contains all the correlations. Hence, this approach is not suitable for cases where the  system and environment states are nonnegligibly correlated initially and/or during the dynamics. The correlated projection superoperator technique\cite{BGM06,B07} improves this situation by mapping the total density matrix to a correlated system-environment state. In this case, the irrelevant part contains less correlation. Therefore, this approach is naturally adapted to cases where the  system has nonnegligible correlation with the environment. Moreover, this method provides accurate results in practical calculations. 

As we will show in this article, the correlated projection superoperator technique has limitations. Its relevant part cannot contain any discord; hence, it is not suitable for cases where the  bipartite  state contains nonnegligible  discord to the environment. The success of the assignment map in the dynamic map inspired this study. The dynamic map describes the evolution of open systems, and a partial trace leads to ambiguities in the dynamic maps\cite{BDMRR13}. A projection is a many-to-one mapping that does not have inverse mappings. The unitary evolution of the total system cannot induce a rigorous map of subsystems, but the assignment map can solve this problem and define an explicitly and rigorously dynamic map.  The projection-operator method in the N-Z equation is similar to the partial trace in dynamic maps. We can replace the projection-operator method with the assignment maps and rewrite the N-Z equation such that the relevant part is obtained directly from the assignment maps. Furthermore, the irrelevant part can be defined as the difference between the density matrix of the total system and the relevant part. In this way, the relevant part can contain discord. The price is the consistent conditions are no longer satisfied\cite{RMA10}. The partial trace of the relevant part cannot return the density matrix of the system. However, the assignment map is a one-to-one mapping: the density matrix of the system can be recovered via inverse mapping of the assignment map. Hence, we can still obtain the dynamics of the system, even though the assignment map is not consistent.  Another  benefit of assignment map  is that the relevant part is determined by the local information of the system; therefore,  the impact of the relevant part is exactly the influence of memory at that period of time.


The benefits of assignment maps are now clear; the next problem is what kinds of assignment maps should be used. Similar to the article\cite{BDMRR13}, we attempt to reduce the correlation in the irrelevant part; hence, we choose an assignment map that maps the density matrix of the system at $t_0$ to the best separable state (BSS) of the total system at $t_0$. The assignment map also has its limitations: the relevant part given by the assignment map is separable state. Therefore, the relevant part cannot  contain entanglement, possibly due to the properties of non-Markovianity, which can be measured with completely positive (CP) maps \cite{RHP10}. We show that the evolution map of the system is CP when there is no entanglement between the system and the environment. Therefore, entanglement is the deciding  factor for non-Markovianity. Additionally, when the irrelevant part vanishes, the evolution should be CP under the weak coupling limit. Hence, the relevant part that contains entanglement may conflict with the previous conclusion. We may not be able to further improve the irrelevant part.


The equivalence of non-Markovianity and information backflow  provides another way to address the issue of non-Markovianity\cite{BLP09}. The propagation of information has been widely discussed in the quantum lattice model\cite{LR72,HK06,HG18}, and many tools to analyze the complex multibody problem in lattice models are available. One of the most important results is the Lieb-Robinson bound, which provides a general upper bound on the information flow. Following this approach, we apply the new N-Z equation in the quantum lattice model to obtain further results. The conclusion is helpful in analyzing  the impact of entanglement.


In the next section, we first show the relation between entanglement flow and non-Markovianity.  We also prove that the usual N-Z equation is not suitable for research on entanglement flow. Then, we rewrite the N-Z equation with assignment maps. In Sec. \ref{NZEITQLM}, we apply the new equation in the quantum lattice model, and we use cluster expansion tools to  bound the inhomogeneous term. With the simplified result, we  analyze properties that affect the non-Markovianity, such as the environment dimensions and coupling strength. Finally, in Sec. \ref{CON}, we briefly discuss  possible improvement of the bound of the inhomogeneous term and conclude the paper.

\section{Why another N-Z equation and how}\label{WANZEAH}
In this section, we first introduce several correlation measures and show that entanglement is the key factor to the CP maps among these correlations. Since the projection method cannot generally recover the separable state, the irrelevant part provided by projection operator techniques inevitably contains excessive unnecessary correlations. The unnecessary correlations are generally not monogamous, which greatly increases the difficulty of computation. In contrast, assignment maps can recover all the correlations except entanglement, which makes the irrelevant part contain fewer correlations. The N-Z equation rewritten via this method can be used in the broader case. Moreover, the monogamy properties of entanglement simplify the calculation of the inhomogeneous term.  The above reason guides us to rewrite the N-Z equation with assignment maps. 
\subsection{Introduce correlation measure}
A general review of correlation can be found in\cite{MPSVW10,S15}. Here, we briefly introduce some quantum and classical correlations, together with the general form of the bipartite density matrix  when these correlations are equal to zero. 

Entanglement is among the most important concepts in quantum information theory. Many measures of entanglement have been proposed\cite{BCY11}. Squashed entanglement satisfies several properties, such as faithfulness, convexity and monogamy\cite{BCY11}.  The equivalence between vanish squashed entanglement and separable states has been proved\cite{CW04}. The separable states can generally be written as $\mathcal{S}_{A:B}=\sum_i P_i \rho_{i}^A\otimes\rho_{i}^B$, where $\rho_{i}$ is the density matrix of the pure state.

Quantum discord is historically the first measure of quantum correlations beyond entanglement and is  generally defined as the difference between the mutual information and the classical correlations. This measure is vanished on classical-quantum or quantum-classical states only\cite{D10}. The classical-quantum states can generally be written as  $\rho_{A:B}=\sum_i P_i \Pi_{i}^A\otimes\rho_{i}^B$, where $\{\Pi_{i}\}$ is the projection operator.

General quantum correlation has been summarized by\cite{ABC16}. If a bipartite state does not have any quantum correlation, it must belong to classically correlated states, which can generally be written as  $\rho_{AB}=\sum_i P_i \Pi_{i}^A\otimes\Pi_{i}^B$.

The classical correlation can be measured by mutual information, which is vanished on  product states $\rho_{AB}= \rho^A\otimes\rho^B$.

\subsection{Monogamy correlation measure beyond entanglement}
A general monogamy correlation measure beyond entanglement does not exist\cite{SAPB12}: the correlation measure should not increase upon attaching a local pure ancilla. The measure should also be invariant under local unitary transformation. Therefore, the  correlation measure should not increase upon Stinespring dilation $\Lambda[\mathcal{S}_{A:B}]=\Tr_C(U_{BC}\mathcal{S}_{A:B}\otimes\ket{0}\bra{0}_CU_{BC}^\dagger)$. For a generic separable state $\mathcal{S}_{A:B}$, the Stinespring dilation can always broadcast $B$ to $C$, which falsifies the monogamy or positivity of the correlation measure.

Squashed entanglement is defined as\cite{BCY11}
\begin{equation}
    E_{sq}(\rho_{A:B})=\frac{1}{2}\min_{\rho\in\{\rho_{ABE}\}} I(A:B|E)_{\rho},
\end{equation}
where $\{\rho_{ABE}\}$ is an extension of $\rho_{AB}$. Squashed entanglement is the only known monogamy entanglement measure and can be used to bound the distance between  $\rho_{AB}$ and its BSS\cite{BCY11}. The BSS is the separable state with the minimum distance to $\rho_{AB}$. We label  the set of separable states on $A:B$ as $\mathcal{S}_{A:B}$. The  minimum distance is  defined as
\begin{equation}\label{bss}
    \norm{\rho_{AB}-\mathcal{S}_{A:B}}=\min_{\sigma\in \mathcal{S}_{A:B}}\norm{\rho_{AB}-\sigma}.
\end{equation}

\subsection{Separable state is the sufficient condition for CP maps}
The system and environment  form an isolated system, and evolution of the isolated system is unitary. The  dynamic map is defined by $\mathcal{B}(\rho_S)=\Tr_E (U\rho_{SE}U^\dagger)$. If there is a CP map $\mathcal{B}$ that satisfies this equation, then the dynamic map is CP.

If the bipartite state has vanished quantum discord, then any unitary transformation of the total system would be CP maps for the system\cite{SL09}. Vanishing quantum discord is not necessary for CP maps\cite{BDMRR13}. The dynamic map $\mathcal{B}$ can be expressed by the assignment map as $\mathcal{B}=\mathcal{M}\circ\mathcal{U}\circ\mathcal{A}$, where $\mathcal{U}$ implies unitary transformation and $\mathcal{M}$ is the trace over the environment. Assignment map $\mathcal{A}$ maps the density matrix of system $\rho_S$ to the density matrix of the total system $\rho_{SE}$. Maps $\mathcal{M}$ and $\mathcal{U}$ are both CP; hence, $\mathcal{B}$ is CP iff assignment map $\mathcal{A}$ is CP.

If the total system is in a separable state, we can prove that there is a linear CP  assignment map that can restore the system from the density matrix. Hence, the separable state is a sufficient condition for CP maps.  The proof is as follows. Separable states can be generally written as $\mathcal{S}_{S:E}=\sum_i P_i \Pi_{\psi_i}^S\otimes\rho_{i}^E$, where $\Pi_{\psi_i}^S=\ket{\psi_i}\bra{\psi_i}$. We can always choose the bases to make the classical-quantum parts $\sum_{i:\Pi_{\psi_i}^S\in \{\Pi_j\}} P_i \Pi_{\psi_i}^S\otimes\rho_{i}^E$ have maximum probability $P_{\text{Max}}=\sup_{ \{\Pi_j\}}\sum_{i:\Pi_{\psi_i}^S\in \{\Pi_j\}} P_i$. Under the bases $ \{\Pi_j\}$, the probability distribution of $S$ is $P_j^S=\sum_i P_i\text{Tr}_S(\Pi_{\psi_i}^S\Pi_j^S)$. The bases $\{\Pi_j\}$ ensure the matrices 
\begin{equation}\label{amm}
    M_{ij}=\sqrt{\frac{P_j}{P_i^S}\text{Tr}(\Pi_{\psi_j}^S\Pi_i^S)}\ket{\psi_j}\bra{i}
\end{equation}
are all finite. With these matrices, the linear CP assignment map 
\begin{equation}\label{am}
    \mathcal{A}\rho_S=\sum_{ij}M_{ij}\rho_S M_{ij}^\dagger\otimes\rho_j^E
\end{equation}
can restore the separable state  $\mathcal{S}_{S:E}=\mathcal{A}\circ\mathcal{M}\mathcal{S}_{S:E}$. It is also easy to check that the completeness relation $\sum_{ij}M_{ij}^\dagger M_{ij}=\mathbbm{1}$ is satisfied. This relation ensures the trace-preserving (TP) property of assignment map $\Tr_{SE}\circ \mathcal{A}=\Tr_{S}$.

Vanishing entanglement is a necessary and sufficient condition for a separable state\cite{CW04}; hence, vanishing entanglement is a sufficient condition for CP maps.

As summarized by \cite{RMA10}, if we want to restore entanglement, the assignment maps are generally linear and consistent but not positive; therefore,  vanishing entanglement may also a necessary condition for CP maps.

\subsection{Projection method only gives zero-discord state}
In this section, we show that the  correlated projection superoperator technique can give only the quantum-classical/classical-quantum states. Hence, in the general situation, the relevant part is not the BSS.

The projection superoperator must satisfy some properties: (i) It is a projection superoperator: $\mathcal{P}^2=\mathcal{P}$. (ii) It is linear: $\mathcal{P}(\alpha\mathcal{O}_1+\beta \mathcal{O}_2)=\alpha \mathcal{P}\mathcal{O}_1+\beta\mathcal{P} \mathcal{O}_2$. (iii)TP: $\Tr\mathcal{P}\mathcal{O}=\Tr\mathcal{O}$.

The general linear projection superoperator $\mathcal{P}$ gives\cite{B07} 
\begin{equation}
    \mathcal{P}\rho=\sum_i \Tr_E\{A_i\rho\}\otimes B_i,
\end{equation}
where $A_i$ and $B_i$ must satisfy $\Tr_E\{B_iA_j\}=\delta_{ij}$ and $\sum_i (\Tr_E B_i)A_i=I_E$. The quantum discord of $\mathcal{P}\rho$ can be measured with\cite{BP02}
\begin{equation}
    D^{E|S}(\mathcal{P}\rho^{SE})=S(E)-S(SE)+\inf_{{M_i^E}}S(S|\{M_i^E\}).
\end{equation}
With ${M_i^E}={A_i}$, it is easy to prove that $S(S|\{A_i\})=S(SE)-S(E)$. Hence, we have $D^{E|S}(\mathcal{P}\rho^{SE})\leq0$.  Since $D^{E|S}$ is always positive, $D^{E|S}(\mathcal{P}\rho^{SE})$ must be zero; thus the  general linear projection superoperator can give only a zero-discord state.


\subsection{ N-Z equation from assignment maps}
The consistent and positive properties of linear assignment maps have been discussed in detail\cite{RMA10}. To restore the quantum correlations, we  need to abandon the positive property. To recover the classical correlations, we need to abandon the consistent property. Here we choose the positive property and recover all the correlations except entanglement. Hence, the assignment map is not generally consistent: it is consistent for only a special state.  We must be careful when handling this component.

From \Eq{\ref{am}}, we can always find an assignment map such that
\begin{equation}\label{bssam}
    \mathcal{A}_{T_0}\circ\mathcal{M} \rho(t_0)=\mathcal{S}_{A:B}(t_0),
\end{equation}
where $\mathcal{S}_{A:B}(t_0)$ is the BSS of $ \rho(t_0)$.
The consistent property for state $ \rho(t_0)$ is naturally satisfied because
\begin{equation}
    \mathcal{M}\circ  \mathcal{A}_{T_0}\circ\mathcal{M} \rho(t_0)=  \mathcal{M}\circ \mathcal{S}_{A:B}(t_0)=\Tr_B \rho(t_0) .
\end{equation}
The consistent property for the general state does not exist: $ \mathcal{M}\circ  \mathcal{A}_{T_0}\circ\mathcal{M}\neq  \mathcal{M}$.

Comparing \Eq{\ref{bssam}} with projection operator techniques, we need only replace $\mathcal{P}$, $\mathcal{Q}$ with $\mathcal{A}_T\circ\mathcal{M}$, $1-\mathcal{A}_T\circ\mathcal{M}$. Therefore, the new N-Z equation is 
\begin{align}\label{nnz}
    \partial_t \mathcal{M}\rho(t)=\mathcal{M}\circ \mathcal{L}(t)\circ\mathcal{G}(t,t_0)(I- \mathcal{A}_{T_0}\circ\mathcal{M})\rho(t_0) \notag\\
    +\int_{t_0}^t ds \mathcal{M}\mathcal{G}(t,s)(I- \mathcal{A}_{T_0}\circ\mathcal{M})\mathcal{L}(s)\mathcal{A}_{T_0}\circ\mathcal{M}\rho(t_s),
\end{align}
where $\mathcal{G}(t,t_0)=T_{\leftarrow}\exp[\int_{t_0}^t ds(I- \mathcal{A}_{T_0}\circ\mathcal{M})\circ\mathcal{L}(s)]$. In projection operator techniques, the super operator satisfies $\mathcal{P}\mathcal{Q}=0$. Here,  the  lack of consistency makes  $(I- \mathcal{A}_{T_0}\circ\mathcal{M})\mathcal{A}_{T_0}\circ\mathcal{M}\rho_{t}\neq0$.  However, we still have $(I- \mathcal{A}_{T_0}\circ\mathcal{M})\mathcal{A}_{T_0}\circ\mathcal{M}\rho_{t_0}=0$.

The RHS of \Eq{\ref{nnz}} corresponds to the influence of the historical state on the present evolution. The integral term in \Eq{\ref{nnz}} is decided by the state of the system within time $[t_0,t]$. Hence, it represents the influence of the historical state within the period $[t_0,t]$. The irrelevant part is nonlocal: it records the history of the system before $t_0$. In open systems, we cannot control or know the whole environment in practice. Moreover, we cannot control or know the whole history of the system. Hence, we must lose the irrelevant part. The more accurate the separation of the local influence is, the less information we lose in practical situations. The irrelevant part 
\begin{equation}
    (I- \mathcal{A}_{T_0}\circ\mathcal{M})\rho(t_0)=\rho(t_0)-\mathcal{S}_{A:B}(t_0)
\end{equation}
contains only entanglement and is smaller than that of the previous method, which also includes discord. Hence, the assignment provides a more accurate evolution. 

\section{N-Z EQUATION of THE QUANTUM LATTICE MODEL} \label{NZEITQLM}
For simplicity,  we first discuss a one-dimensional lattice with
the nearest-neighbor interaction. Then, we briefly consider higher-dimensional models. 

Suppose $C_n:=\{x\in B|d(x,A)=n\}$ are all sites in an environment with distance $n$ from the system. We set $B_n:=\{x\in B|d(x,A)> n\}$ and $\bar{B}_n:=B-B_n$. The trace over the whole environment can be generalized to the partial environment $\mathcal{M}_n=\Tr_{B_n}\circ$. Clearly, 
\begin{equation}\label{ptopt}
    \mathcal{M}_{n\leq m}\mathcal{M}_m=\mathcal{M}_{n}.
\end{equation}
We can also generalize the assignment map operator as $\mathcal{A}_n^{T_0}\circ\mathcal{M}_n\rho(t_0):=\mathcal{S}_{A\bar{B}_n:B_n}(t_0)$, where $\mathcal{S}_{A\bar{B}_n:B_n}(t_0)$ is the BSS of $A\bar{B}_n:B_n$ at time $t_0$. 
$(I-\mathcal{A}_n^{T_0}\circ\mathcal{M})\rho(t_0)=\rho(t_0)-\mathcal{S}_{A\bar{B}_n:B_n}(t_0)$  reflects the entanglement between the $B_n$ and $A\bar{B}_n$. According to the definition, $\mathcal{M}_n\circ\mathcal{A}_n^{T_0}\circ\mathcal{M}_n\rho(t_0)=\rho_{A\bar{B}_n}(t_0)=\mathcal{M}_n\rho(t_0)$.
It is easy to prove
\begin{align}\label{project}
    \mathcal{A}_m^{T_0}\circ\mathcal{M}_m\circ\mathcal{A}_{n\geq m}^{T_0}\circ\mathcal{M}_{n\geq m}\rho(t_0)=\mathcal{A}_m^{T_0}\circ\mathcal{M}_m \rho(t_0)\notag\\
    \mathcal{A}_{m\leq n}^{T_0}\circ\mathcal{M}_{m\leq n}(I-\mathcal{A}_n^{T_0}\circ\mathcal{M}_n)\rho(t_0)=0.
\end{align}
In the case of the nearest-neighbor interaction,  the total Hamiltonian can be generally written as
\begin{equation}
    H_{tot}=H_A+H_{AC_1}+H_{C_1}+H_{C_1C_2}+H_{C_2}+\dots,
\end{equation}
where $H_{C_i}$ describes the local Hamiltonian of a single site and $H_{C_iC_{i+1}}$ describes the interaction among sites. We set $H_0=H_A+\sum_{i}H_{C_i}$ and $H_I=H_{AC_1}+\sum_{i}H_{C_iC_{i+1}}$. Under the interaction representation,  the equation of motion of the density matrix is
\begin{equation}
    \partial_t \rho(t)=-i[H_I(t),\rho(t)]\equiv\mathcal{L}(t)\rho(t),
\end{equation}
where $H_I(t)=\exp(iH_0t)H_I\exp(-iH_0t)$. We can separate the interaction into three parts $H_I^n=H_{AC_1}+\sum_{i<n}H_{C_iC_{i+1}}$, $H_I^n=H_{C_nC_{n+1}}$ and $H_I^{\bar{n}}=\sum_{i>n}H_{C_iC_{i+1}}$. Their corresponding Liouville superoperators are $\mathcal{L}_{n}(t)$, $\mathcal{L}_{n,n+1}(t)$ and  $\mathcal{L}_{\bar{n}}(t)$. It is easy to prove that
\begin{equation}\label{projectl}
    \mathcal{M}_n \mathcal{L}_{\bar{n}}(t)=0, \mathcal{M}_n \mathcal{L}_{n}(t)= \mathcal{L}_{n}(t) \mathcal{M}_n.
\end{equation}
These properties are useful in simplifying the N-Z equation. For example, we have
\begin{widetext}
\begin{align*}
    \mathcal{A}_0^{T_0}\circ\mathcal{M}_0 \mathcal{L}_{tot}(t)(I- \mathcal{A}_1^{T_0}\circ\mathcal{M}_1)\rho(t_0)  = \mathcal{A}_0^{T_0}\circ\mathcal{M}_0(  \mathcal{L}_{AC_1}(t))(I- \mathcal{A}_1^{T_0}\circ\mathcal{M}_1)\rho(t_0)\\
    = \mathcal{A}_0^{T_0}\circ\mathcal{M}_0\circ\mathcal{M}_1\mathcal{L}_{AC_1}(t)(I- \mathcal{A}_1^{T_0}\circ\mathcal{M}_1)\rho(t_0)= \mathcal{A}_0^{T_0}\circ\mathcal{M}_0 \mathcal{L}_{AC_1}(t)\circ\mathcal{M}_1(I- \mathcal{A}_1^{T_0}\mathcal{M}_1)\rho(t_0) =0.
    \end{align*}
$(I- \mathcal{A}_1^{T_0}\mathcal{M}_1)\rho(t_0) $ corresponds to the information spread over $B_1$. Through a single interaction, this part of information cannot flow to $A$. Thus, it cannot impact $A$ and $  \mathcal{A}_0^{T_0}\circ\mathcal{M}_0 \mathcal{L}_{tot}(t)(I- \mathcal{A}_1^{T_0}\circ\mathcal{M}_1)\rho(t_0)=0$. In addition, with \cref{ptopt,projectl}, we can  give the bound of the partial trace of the commutator of interaction
\begin{equation}\label{bptcio}
    \norm{ \mathcal{M}_{k}\mathcal{L}_{n-1,n}(t) \mathcal{O}}_1=  \norm{ \mathcal{M}_{k}\mathcal{L}_{n-1,n}(t)   \mathcal{M}_{\Max\{k,n\}}\mathcal{O}}_1 
    \leq \norm{ \mathcal{L}_{n-1,n}(t)   \mathcal{M}_{\Max\{k,n\}}\mathcal{O}}_1 
    \leq 2\norm{ H_{n-1,n}}_\infty \norm{   \mathcal{M}_{\Max\{k,n\}}\mathcal{O}}_1.
\end{equation}

The dynamics of the system can be described by 
\begin{equation}\label{P}
    \partial_t   \mathcal{M}_0 \rho(t)=\mathcal{M}_0\circ \mathcal{L}_{1}(t)  \rho(t)
    =\mathcal{M}_0  \mathcal{L}_{1}(t)\mathcal{M}_1  (I- \mathcal{A}_0^{T_0}\circ\mathcal{M}_0) \rho(t),
 \end{equation}
where $\mathcal{L}_1(t)\equiv\mathcal{L}_{AC_1}(t)$.
We  repeatedly use \Eq{\ref{project}} and  \Eq{\ref{projectl}} to derive this equation. Similarly, the evolution of the irrelevant part is from
\begin{equation}\label{Q}
    \partial_t(I- \mathcal{A}_0^{T_0}\circ\mathcal{M}_0) \rho(t)=( (I- \mathcal{A}_0^{T_0}\circ\mathcal{M}_0) \mathcal{L}_{1}(t)+ \mathcal{L}_{\bar{0}}(t))\mathcal{A}_0^{T_0}\circ\mathcal{M}_0 \rho(t)+( (I- \mathcal{A}_0^{T_0}\circ\mathcal{M}_0)\mathcal{L}_{1}(t)+ \mathcal{L}_{\bar{0}}(t))(I- \mathcal{A}_0^{T_0}\circ\mathcal{M}_0) \rho(t).
\end{equation}
Given $\rho(t_0)$ at initial time $t_0$, the formal solution of \Eq{\ref{Q}}  is
\begin{equation}\label{fQ}
    (I- \mathcal{A}_0^{T_0}\circ\mathcal{M}_0) \rho(t)=\mathcal{G}(t,t_0)(I- \mathcal{A}_0^{T_0}\circ\mathcal{M}_0) \rho(t_0)+\int_{t_0}^t ds \mathcal{G}(t,s)( \mathcal{L}_1'(t)+ \mathcal{L}_{\bar{0}}(t))\mathcal{A}_0^{T_0}\circ\mathcal{M}_0 \rho(s),
\end{equation}
where $\mathcal{G}(t,t_0)=\mathcal{T}_{\leftarrow}\exp(\int_{t_0}^t ds(\mathcal{L}_1'(s)+ \mathcal{L}_{\bar{0}}(s)))$ and $\mathcal{L}_1'(t)= (I- \mathcal{A}_0^{T_0}\circ\mathcal{M}_0) \mathcal{L}_{1}(t)$. Inserting \Eq{\ref{fQ}} into \Eq{\ref{P}}, we obtain 
\begin{equation}\label{nz}
    \partial_t \mathcal{M}_0 \rho(t)=\mathcal{M}_0 \mathcal{L}_{1}(t) \mathcal{G}(t,t_0)\Delta(t_0)
    +  \int_{t_0}^t ds\mathcal{M}_0 \mathcal{L}_{1}(t)\mathcal{G}(t,s)(\mathcal{L}_1'(t)+ \mathcal{L}_{\bar{0}}(t))\mathcal{A}_0^{T_0}\circ\mathcal{M}_0 \rho(s),
 \end{equation}
where $\Delta(t_0)=\rho(t_0)-\mathcal{S}_{A:B}(t_0) $.

In the following text, we bound the inhomogeneous term. Before doing so, we show that the properties of superoperator $\mathcal{L}_1'(t)$ are very similar to those of $\mathcal{L}_1(t)$. First, the partial trace $ \mathcal{M}_{d>0} $ is also commutative with $\mathcal{L}_1'(t)$,  which is similar to \Eq{\ref{projectl}}
\begin{equation}\label{projectll}
    \mathcal{M}_{d>0} \mathcal{L}_1'(t)= \mathcal{M}_{d} \mathcal{L}_1'(t)\mathcal{M}_{d}.
\end{equation}
Second, with the definition \Eq{\ref{amm}}, it is easy to prove that the assignment map cannot increase the trace class norm of any operator
\begin{equation}\label{asn}
    \norm{\mathcal{A}_0^{T_0}\circ\mathcal{M}_0\mathcal{O}}_1=  \norm{\sum_{ij}\frac{P_j}{P_i^S}\Tr(\Pi_{\psi_j}^S\Pi_i^S)\ket{\psi_j}\bra{\psi_j}\mathcal{O}^A_{ii}}_1
    \leq \sum_{ij}\frac{P_j}{P_i^S}\Tr(\Pi_{\psi_j}^S\Pi_i^S)\abs{\mathcal{O}^A_{ii}}\times\norm{\ket{\psi_j}\bra{\psi_j}}_1  
    =\sum_{i}\abs{\mathcal{O}^A_{ii}}\leq  \norm{\mathcal{O}}_1,
\end{equation}
where $\mathcal{O}^A=\mathcal{M}_0\mathcal{O}$. This property, together with \Eq{\ref{projectll}}, leads to a similar bound as that in \Eq{\ref{bptcio}}
\begin{equation}\label{bptciol}
    \norm{ \mathcal{M}_{k}\mathcal{L}'_1(t) \mathcal{O}}_1\leq 2 \norm{\mathcal{L}'_1(t)  \mathcal{M}_{k}\mathcal{O}}_1\leq 4\norm{ H_{AC_1}}_\infty \norm{   \mathcal{M}_{k}\mathcal{O}}_1.
\end{equation}
The superoperator $ I- \mathcal{A}_0^{T_0}\circ\mathcal{M}_0$ can double the bound at most. Comparing \cref{projectl,bptcio} with \cref{projectll,bptciol}, we conclude that the operator $\mathcal{L}'_1(t)$ can  be treated the same as the other superoperators in $ \mathcal{L}_{\bar{0}}(t)$. The only difference is that the interaction strength should be doubled when we make the bound.

Now, we start to determinethe bound of the inhomogeneous term. We adopt a similar approach to that  used in the Lieb-Robinson bound\cite{HK06}. The main idea is that the information flow back to the system is limited in the short term.  Making use of  \cref{projectl,projectll}, we have
    \begin{align}
       \norm{ \mathcal{M}_0 \mathcal{L}_{1}(t) \mathcal{G}(t,t_0)\Delta(t_0) }_1=\norm{ \mathcal{M}_0 \mathcal{L}_{1}(t) \mathcal{M}_1e^{\sum_{Z_1}\Delta t \mathcal{L}_{Z_1}(t-\Delta t)}\mathcal{G}(t-\Delta t,t_0)\Delta(t_0) }_1 \notag \\
     \leq  \norm{ \mathcal{M}_0 \mathcal{L}_{1}(t-\Delta t) \mathcal{G}(t-\Delta t,t_0)\Delta(t_0) }_1 + \norm{ \mathcal{M}_0 \mathcal{L}_{1}(t)\sum_{Z_1}\Delta t \mathcal{L}_{Z_1}(t-\Delta t) \mathcal{G}(t-\Delta t,t_0)\Delta(t_0) }_1 \notag \\
     \leq  \norm{ \mathcal{M}_0 \mathcal{L}_{1}(t-\Delta t)\mathcal{G}(t-\Delta t,t_0) \Delta(t_0)}_1 + \sum_{Z_1}2\Delta t \norm{ H_{AC_1}}_\infty \norm{\mathcal{M}_1 \mathcal{L}_{Z_1}(t-\Delta t)\mathcal{G}(t-\Delta t,t_0)\Delta(t_0) }_1,
    \end{align}
where $Z_1:Z_1\cap{AC_1}\neq0$. Using the above bound iteratively and letting $\Delta t\to 0$, we obtain
    \begin{equation}\label{ib}
        \norm{ \mathcal{M}_0 \mathcal{L}_{1}(t)\mathcal{G}(t,t_0)\Delta(t_0) }_1\leq \norm{ \mathcal{M}_0 \mathcal{L}_{1}(t_0) \Delta(t_0)}_1  + \sum_{Z_1:Z_1\cap{AC_1}\neq0}2\int_{t_0}^{t} ds\norm{ H_{AC_1}}_\infty \norm{\mathcal{M}_1 \mathcal{L}_{Z_1}(s)\mathcal{G}(s,t_0)\Delta(t_0) }_1.
    \end{equation}
    Making use of \cref{projectl,projectll} again, we can obtain the following general bound:
    \begin{align} \label{gb}
        \norm{ \mathcal{M}_{d\geq d(Z_k)-1} \mathcal{L}_{Z_k}(s)\mathcal{G}(s,t_0)\Delta(t_0) }_1\notag \\
        = \norm{ \mathcal{M}_{d\geq d(Z_k)-1} e^{-\Delta t \mathcal{L}'_{d}(s-\Delta t)}\mathcal{L}_{Z_k}(s) \mathcal{M}_{d'=\Max\{d,d(Z_k)\}}e^{\sum_{Z_{k+1}}\Delta t \mathcal{L}_{Z_{k+1}}(s-\Delta t)} \mathcal{G}(s-\Delta t,t_0)\Delta(t_0)}_1 \notag \\
        \leq 2\norm{  H_{Z_k}}_\infty \norm{\mathcal{M}_{d'=\Max\{d,d(Z_k)\}}e^{\sum_{Z_{k+1}}\Delta t \mathcal{L}_{Z_{k+1}}(s-\Delta t)} \mathcal{G}(s-\Delta t,t_0)\Delta(t_0)}_1,
    \end{align}
   where  $Z_{k+1}:Z_{k+1}\cap{Z_{k}}\neq0\text{ or }  Z_{k+1}=(d,d+1)$. $\mathcal{L}'_{d}$ includes all the interactions in $\mathcal{L}_{d}$ except for the terms that are already included in $\{Z_{k+1}\}$. $ d(Z_k)$ is the minimum distance that satisfies $Z_k\subset\bar{B}_{d(Z_k)}$. The reason we can move the terms about $\mathcal{L}'_{d}$ to the front of $\mathcal{L}_{Z_k}(s)$ is that they are communicable to $\mathcal{M}_{d}$, and the Schatten 1-norm is invariant under unitary transformation. For example, if unitary transformation $U$ is not related to the Hilbert space of ${B_d}$, then  $ \norm{\Tr_{B_d}[H_{Z_{k}},U \mathcal{O}U^\dagger] }_1= \norm{U^\dagger\Tr_{B_d}[H_{Z_{k}},U \mathcal{O}U^\dagger] U}_1= \norm{\Tr_{B_d}[U^\dagger H_{Z_{k}}U,\mathcal{O}] }_1$. Because  $Z_{k+1}$ must overlap  $Z_{k}$ or $\bar{B}_d$, it is easy to verify that  $d'=\Max\{d,d(Z_k)\}\geq d(Z_{k+1})-1$. 
Using the bound \Eq{\ref{gb}} iteratively in \Eq{\ref{ib}}, we obtain
    \begin{equation}\label{flbn}
        \norm{ \mathcal{M}_0 \mathcal{L}_{1}(t) \mathcal{G}(t,t_0)\Delta(t_0)}_1\leq \sum_{k=0}^{\infty}\sum_{g_k} \frac{2^{k+1}(t-t_0)^k}{k!}\norm{ H_{g_k}}_\infty \norm{\mathcal{M}_{d(g_k)}\Delta(t_0) }_1,
    \end{equation}
where $g_k$ is marked by a sequence of bonds $(l_0,l_1,l_2, \dots ,l_k)$, $l_0=AC_1$ and $\norm{ H_{g_k}}_\infty=\prod_{l\in {g_k}}\norm{ H_{l}}_\infty$.  The sequence of bonds can be regarded as a graph. For the nontrivial graph, each bond must overlap the previous bonds $ l_{i+1}\cap l_i\neq0$ or the boundary of the previous graph $ l_{i+1}\cap d(l_0,l_1,l_2, \dots ,l_i)\neq 0 $. $d(g_k)$ is the minimum distance that satisfies $g_k\subset\bar{B}_{d(Z_k)}$. 
Clearly, $d(g_k)\leq k+1$; hence, 
\begin{equation}\label{eblr}
    \sum_{k=0}^{\infty}\sum_{g_k} \frac{2^{k+1}(t-t_0)^k}{k!}\norm{ H_{g_k}}_\infty \norm{\mathcal{M}_{d(g_k)}\Delta(t_0)}_1
    \leq  \sum_{d=1}^{\infty} \sum_{k=d-1}^{\infty}\sum_{g_{k,d}} \frac{2^{k+1}(t-t_0)^k}{k!}\norm{ H_{g_{k,d}}}_\infty \norm{\mathcal{M}_{d}\Delta(t_0) }_1.
\end{equation}
\end{widetext}
The term $\norm{\mathcal{M}_{d}\Delta(t_0) }_1$ is related to the information within distance $d$. Hence, only through $k\geq d$ steps of interaction can all this information influence the evolution of the system. $g_{k,d}$ is all  the possible $k$ bond graphs that satisfy the previous requirement of connectivity and $d(g_{k,d})=d$. The number of such  graphs is less than $4^k$: three possible bonds overlap the last bond and one  bond overlaps the boundary of the graph.  When the interaction strength is approximately the same, we can bound the influence as
\begin{align}\label{LR}
    \sum_{k=d-1}^{\infty}\sum_{g_{k,d}} \frac{2^{k+1}(t-t_0)^k}{k!}\norm{ H_{g_k}}_\infty \notag\\
    \leq \sum_{k=d-1}^{\infty} \frac{2^{k}(t-t_0)^k 4^kJ^{k+1}}{k!}
    \leq J \times y(t-t_0,d-1) ,
\end{align}
where $y(x,L)=e^{8xJ}(8xJ)^L\gamma^*(L,8xJ)$. $\gamma^*$ is the holomorphic extension of the lower incomplete gamma function. $J=\Max\{\sup_Z \norm{H_Z}_\infty, 2\norm{H_{AC_1}}_\infty\}$.  Using the bound of  Stirling's approximation, it is straightforward to obtain 
\begin{equation}
    \frac{1}{(L+k)!}\leq \frac{e^L}{k!L^k}.
\end{equation}
With this relation,  it is easy to prove  $y(x,L)\leq e^{-\mu(x,L) L+vx}$, where $v=8J$ and $\mu(x,L)=-\ln (8eJx/L)$. The property that the influence exponentially decays over distance  is very similar to the  Lieb-Robinson bound. The further away  from the system, the less information influences the evolution.

Now, we  explore the bound of $ \mathcal{M}_d( \rho(t_0)-\mathcal{S}_{A:B}(t_0)) $. $ \rho(t_0)-\mathcal{S}_{A:B}(t_0)$ is decided by the entanglement between $A$ and $B$, and its partial trace is related to  the entanglement between $A$ and $\bar{B}_d$. The entanglement is monogamous if the environment is permutation invariant and sufficiently large. The entanglement between $A$ and $\bar{B}_d$ should be negligible, which means $ \mathcal{M}_d( \rho(t_0)-\mathcal{S}_{A:B}(t_0)) $ should also be negligible for small $d$. We prove this property in detail in the following text.

The general form of $\rho$ is
\begin{equation} \label{gfor}
    \rho=P^0\rho^{tri}_{A\bar{B}_dB_d}+P^1\mathcal{S}_{A:B}+P^2\mathcal{S}_{AB_d:\bar{B}_d}+P^3\mathcal{S}_{A\bar{B}_d:B_d},
\end{equation}
where $\rho^{tri}_{A\bar{B}_dB_d}$ is a three-body entangled state. If there is no entanglement between $AB_d$ and $\bar{B}_d$, the density matrix is
\begin{equation}
    \rho'=(1-P_2)\mathcal{S}_{A:B}+P^2\mathcal{S}_{AB_d:\bar{B}_d}.
\end{equation}
Suppose the $A:B$ BSS of $\rho'$ is $\mathcal{S}'_{A:B}$. In general, the BSS of a mixed quantum state cannot be obtained directly from the BSS of each pure state, but, as we are going to prove, the separable state $\sigma=(1-P_2)\mathcal{S}_{A:B}+P^2\mathcal{S}^2_{A:B}$, where $\mathcal{S}^2_{A:B}$ is the  $A:B$ BSS of $\mathcal{S}_{AB_d:\bar{B}_d}$, is  just $\mathcal{S}'_{A:B}$. The proof is as follows. On the one hand, based on the definition of BSS, we have $\norm{\rho'-\mathcal{S}'_{A:B}}_1\leq\norm{\rho'-\sigma}_1=P^2\norm{\mathcal{S}_{AB_d:\bar{B}_d}-\mathcal{S}^2_{A:B}}$. On the other hand, the distance should not increase when we drop some parts of the state. This leads $\norm{\rho'-\mathcal{S}'_{A:B}}\geq P^2\norm{\mathcal{S}_{AB_d:\bar{B}_d}-\mathcal{S}^2_{A:B}}$. Based on these two facts, we conclude that 
\begin{equation}
    \mathcal{S}'_{A:B}=(1-P_2)\mathcal{S}_{A:B}+P^2\mathcal{S}^2_{A:B}.
\end{equation}
The state that contains minimal entanglement between $AB_d$ and $\bar{B}_d$ can be written as
\begin{equation}
    \rho=\rho'+\epsilon(\Delta \rho^0+\Delta \rho^3),
\end{equation}
where $\Delta \rho^0=P^0(\rho^{tri}_{A\bar{B}_dB_d}-\mathcal{S}^0_{A:\bar{B}_d:B_d})$ and $\Delta \rho^3=P^3(\mathcal{S}_{A\bar{B}_d:B_d}-\mathcal{S}^3_{A:\bar{B}_d:B_d})$. $\epsilon$ is a small number. $\mathcal{S}^0_{A:\bar{B}_d:B_d}$ is the ${A:\bar{B}_d:B_d}$ BSS of $\rho^{tri}_{A\bar{B}_dB_d}$. Similar to \cref{bssam}, it must satisfy
\begin{equation}\label{lbss0}
    \mathcal{M}_d \mathcal{S}^0_{A:\bar{B}_d:B_d}=\mathcal{S}^0_{A:\bar{B}_d},
\end{equation}
where $\mathcal{S}^0_{A:\bar{B}_d}$ is the $A:\bar{B}_d$ BSS of $\mathcal{M}_d \rho^{tri}_{A\bar{B}_dB_d}$. $\mathcal{S}^3_{A:\bar{B}_d:B_d}$ is the ${A:\bar{B}_d:B_d}$  BSS of $\mathcal{S}_{A\bar{B}_d:B_d}$, which must also satisfy
\begin{equation}\label{lbss3}
    \mathcal{M}_d \mathcal{S}^3_{A:\bar{B}_d:B_d}=\mathcal{S}^3_{A:\bar{B}_d},
\end{equation}
where $\mathcal{S}^3_{A:\bar{B}_d}$ is the $A:\bar{B}_d$ BSS of $\mathcal{S}_{A\bar{B}_d:B_d}$.
The  BSS $\mathcal{S}_{A:B}$ can be regarded as a function of the density matrix $\rho$. Then,  the expansion of this function around $\rho'$ is
\begin{equation}\label{eosar}
    \mathcal{S}_{A:B}=f(\rho)=\mathcal{S}'_{A:B}+\epsilon(\Delta \rho^0+\Delta \rho^3)f'(\rho')+o(\epsilon^2).
\end{equation}
$\mathcal{S}_{AB_d:\bar{B}_d}$ can be generally written as $\sum P_i \rho_{AB_d}^i\otimes \rho_{\bar{B}_d}^i$, and its corresponding $A:B$ BSS is  $\sum P_i \mathcal{S}_{A:B_d}^i\otimes \rho_{\bar{B}_d}^i$, where $ \mathcal{S}_{A:B_d}^i$ is $A:B_d$ BSS of $ \rho_{AB_d}^i$. Clearly, $\mathcal{M}_d\mathcal{S}_{AB_d:\bar{B}_d}=\mathcal{M}_d \mathcal{S}^2_{A:B}$. Combining this result with \cref{eosar},  we obtain
\begin{equation}
    \norm{\mathcal{M}_d(\rho- \mathcal{S}_{A:B})}_1\leq\epsilon\norm{\mathcal{M}_d(\Delta \rho^0+\Delta \rho^3)(1-f')}_1+o(\epsilon^2).
\end{equation}
We assume that the function $f$ is smooth around $\rho'$. Then, we can obtain 
\begin{align}\label{GTLE}
   \norm{\mathcal{M}_d(\Delta \rho^0+\Delta \rho^3)(1-f')}_1\leq \notag\\
   \norm{1-f'}_\infty  \norm{\mathcal{M}_d(\Delta \rho^0+\Delta \rho^3)}_1\notag  \leq\\
 C\norm{\mathcal{M}_d(\Delta \rho^0+\Delta \rho^3)}_1    \leq C\norm{\mathcal{M}_d\Delta \rho^0}_1+C\norm{\mathcal{M}_d\Delta \rho^3}_1.
\end{align}
The general three-body entangled state can be written as $\rho^{tri}_{A\bar{B}_dB_d}=\sum_j  P_j \Pi^{tri}_{\tau_j}$, where $\ket{\tau_j}=\sum_i \alpha_{ij} \ket{\psi_{ij}^A\phi_{ij}^{\bar{B}_d}\xi_{ij}^{B_d}}$. The local density matrix is the separable state  
\begin{equation}\label{tris}
    \mathcal{M}_d\rho^{tri}_{A\bar{B}_dB_d}=\sum_{ij} |\alpha_{ij}|^2 \Pi_{\psi_{ij}}^A\otimes \Pi_{\phi_{ij}}^{\bar{B}_d}\in \mathcal{S}^0_{A:\bar{B}_d}.
\end{equation}
Therefore, from \cref{lbss0,tris},  we obtain $\norm{\mathcal{M}_d\Delta \rho^0}_1=0$. 
From  \Eq{\ref{lbss3}}, we obtain $\norm{\mathcal{M}_d\Delta \rho^3}_1=P^3\norm{\mathcal{M}_d\mathcal{S}_{A\bar{B}_d:B_d}-\mathcal{S}^3_{A:{B}_d}}_1\leq\norm{\mathcal{M}_d\rho-\mathcal{S}_{A:\bar{B}_d}}_1$, which is simply the distance between the state of subsystem $\rho_{A\bar{B}_d}$ and its BSS. This distance can be bounded by the squashed entanglement between $A$ and $\bar{B}_d$\cite{BCY11}. 
If bipartite state $\rho_{A:\bar{B}_d}$ is extendible to the overall environment, $B=\{B_1,\dots,B_k\}$ and $\rho_{A:B_1,\dots,B_k}$  is permutation-symmetric in the $B$ systems $\rho_{A:\bar{B}_d}=\Tr_{B_2,\dots,B_k}(\rho_{A:B_1,\dots,B_k})$, then the information should be  distributed evenly over the environment. In this case, the distance can be bounded as \cite{BCY11}
\begin{equation}\label{DFSE}
    \norm{\rho_{A\bar{B}_d}-\mathcal{S}_{A:\bar{B}_d}}_1\leq\sqrt{918\ln2\times\abs{A}\log\abs{A}}\times \sqrt{\frac{\abs{\bar{B}_d}}{k}},
\end{equation}
where $\abs{A}$ is the freedom of system $A$. The freedom of $\bar{B}_d$ increases exponentially with the number of sites in $\bar{B}_d$ and generally approximates to  $\exp(\beta d)$, where $\beta$ is determined by the dimension of the Hilbert space of a single site.  $k$ is inversely proportional to the  number of sites in $\bar{B}_d$ and approximates to  $L/d$, where $L$ is the length scale of environment $B$. The bound of $\norm{\rho_{A\bar{B}_d}-\mathcal{S}_{A:\bar{B}_d}}_1$ increases exponentially with $d$. When $d$ is  sufficiently large, this bound is worse than the trivial bound $ \norm{\rho_{A\bar{B}_d}-\mathcal{S}_{A:\bar{B}_d}}_1\leq1/2$. We denote this distance as $D$. 

Combining  \cref{nz,flbn,eblr,LR,GTLE,DFSE}, we can bound the inhomogeneous term as
\begin{align}\label{btit}
    \norm{\mathcal{M}_0 \mathcal{L}_{0}\mathcal{G}(t,t_0)\Delta(t_0) }_1\leq \notag\\
   \sum_{ d}\sum_{ g\in G_d} \frac{2^{|g|+1}(t-t_0)^{|g|}}{|g|!} \norm{H_{g}}_\infty \norm{\mathcal{M}_d\Delta(t_0)}_1  \leq \notag  \\
 \sum_{d\leq D}C''*d^{1/2}e^{-(\mu-\beta/2) d+v(t-t_0)}/L^{1/2} \notag\\
 +\sum_{d> D}C'*e^{-\mu d+v(t-t_0)} ,
\end{align}
where $C'=C*J/2$, $C''=C*J*\sqrt{918\ln2\times\abs{A}\log\abs{A}}$.
When the length scale of the environment is large,  the $k$ in \cref{DFSE} is large, which means $\mathcal{M}_d(\rho- \mathcal{S}_{A:B})$ is very small for small $d$. Therefore, the terms in the third line of \cref{btit} are very small. For large $d$, the Lieb-Robinson bound limits the impact of information flow. The terms in the fourth line of \cref{btit} are also be  very small. In conclusion, the inhomogeneous term is negligible when the environment is sufficiently large, $t-t_0$ is sufficiently short and  $\rho_{A:\bar{B}_d}$  is extendible to the overall environment for any $d$.

In contrast to the inhomogeneous term, the memory kernel term is not bounded by the length scale of the environment but is highly dependent on $H_I$. Only  higher-order interactions, such as $O(H_I^4)$ or beyond, can introduce non-Markovian effects. Hence, in some strong coupling limits, non-Markovian behavior always exists,  independently of the environment dimensions\cite{ZPG11}.

Now, we briefly discuss how to bound the inhomogeneous term in higher-dimensional lattices. The \cref{nz} is still applicable, and we  use $ \mathcal{M}_{g_{n}}=\Tr_{g_{n}}\circ$ to  help us simplify the  inhomogeneous term,  where $g_{n}$ corresponds to all the sites in the graph $g_{n}$. Similar to \cref{bptcio}, with \cref{projectl,projectll}, we have
\begin{align}
    \norm{ \mathcal{M}_{g_n}\mathcal{L}_{Z_n} \mathcal{O}}_1&=  \norm{ \mathcal{M}_{g_n}\mathcal{L}_{Z_n}  \mathcal{M}_{g_{n+1}}\mathcal{O}}_1 \notag\\
    &\leq 2\norm{ {H}_{Z_n}}_\infty \norm{ \mathcal{M}_{g_{n+1}}\mathcal{O}}_1,
\end{align}
where $g_{n+1}$ is the union graph of $g_{n}$ and $ {Z_n}$.
Similar to \cref{gb}, we have
\begin{align} 
    \norm{ \mathcal{M}_{g_{k-1}} \mathcal{L}_{Z_k}(s)\mathcal{G}(s,t_0)\Delta(t_0) }_1\notag \\
    = \norm{\mathcal{M}_{g_{k-1}}\mathcal{L}_{Z_k}(s) \mathcal{M}_{g_{k}} \mathcal{U}(\Delta t )\mathcal{G}(s-\Delta t,t_0)\Delta(t_0)}_1,
\end{align}
where $g_k=g_{k-1}\cup Z_k$ and $\mathcal{U}(\Delta t )=e^{\sum_{Z_{k+1}}\Delta t \mathcal{L}_{Z_{k+1}}(s-\Delta t)}$. If $Z_{k+1}$ does not connect $g_k$, it is eliminated by $\mathcal{M}_{g_k}$. If $Z_{k+1}$ does  connect $g_k$ but belongs to $g_{k-1}$, then it is communicable to $\mathcal{M}_{g_{k-1}}$. In this case, if it does not connect $Z_{k}$, then it will be eliminated by the norm. In conclusion, only the interactions that connect $Z_{k}$ or connect $g_{k}$ but do not belong to $g_{k-1}$ are nontrivial, i.e.,   ${Z_{k+1}:Z_{k+1}\cap{Z_{k}}\neq0\lor \{Z_{k+1}\cap g_{k-1}\neq0 \land Z_{k+1}\notin g_{k-1}}\}$. Counting such graphs is not easy.  In this article, we simplify the problem to counting the graphs that connect $A$ without any isolated blocks. The set of such graphs  includes the previous set, but, as we will see below, the bound is still meaningful even when using such a loose constraint.

Suppose the sequence of bonds $(l_0,l_1,l_2, \dots ,l_{k-1})$ contains $i$ nonrepetitive bonds; then, the $i$ nonrepetitive bonds form an animal\cite{HG18} that connects $A$. The $i$ bond animals can occupy at most $i$  sites in $B$. In this way, similar to \cref{flbn}, we have
\begin{align}
    \norm{\mathcal{M}_0 \mathcal{L}_{0}\mathcal{G}(t,t_0)\Delta(t_0) }_1\leq \notag\\
   \sum_{ i}\sum_{ k=i}^{\infty}\sum_{ g\in G_{k,i}} \frac{2^{k+1}(t-t_0)^{k}}{k!} \norm{H_{g}}_\infty \norm{\mathcal{M}_i\Delta(t_0)}_1,
\end{align}
where $ G_{k,i}$ is all the possible $k$ bond graphs with $i$ nonrepetitive bonds. 

The number of possible $i$ bond animals connected to $A$ is less than $|l_0|(Ke)^i$\cite{MS11}, where $|A|$ is the number of possible $l_0$,  $K$ is the valence of each site and $e$ is the natural constant. The possible sort order of $i$ nonrepetitive bonds is less than $i!$. The possible choices that select $i$  nonrepetitive bonds from $k$ bonds is $C^k_{i}$. The possible choices for the remaining $k-i$ repetitive bonds is less than $i^{k-i}$.  In conclusion, the number of graphs in $G_{k,i}$  is less than $|l_0|(Ke)^ii!C^k_{i}i^{k-i}$. Thus, we obtain
\begin{align}
  \sum_{ k=L}^{\infty}\sum_{ g\in G_{k,L}} \frac{2^{k+1}(t-t_0)^{k}}{k!} \norm{H_{g}}_\infty\notag \\
   \leq \sum_{ k=L}^{\infty} \frac{2|l_0|(2KeJ(t-t_0))^L (2LJ(t-t_0))^{k-L}} {(k-L)!} \notag \\
   \leq 2|l_0|e^{-\mu' L+v'(t-t_0)}.
\end{align}
where $v'=2LJ$, $\mu'=-\ln (2KeJ(t-t_0))$.
The influence still decays exponentially over distance,  but the decay rate is smaller. Moreover, the rate of increase over time is also  faster. 

Following  an argument similar to that of Eqs. (\ref{gfor}) to (\ref{DFSE}), the bound of $\norm{\mathcal{M}_d\Delta(t_0)}_1$  is proportional to  $(d\exp(\beta d)/L^\alpha)^{1/2}$, where $L^\alpha$ is equal to the number of sites in $B$.

Similar to \cref{btit},  the inhomogeneous term can be bounded as
\begin{align}
    \norm{\mathcal{M}_0 \mathcal{L}_{0}\mathcal{G}(t,t_0)\Delta(t_0) }_1\leq \notag\\
 \sum_{d\leq D}C''*d^{1/2}e^{-(\mu'-\beta/2) d+v'(t-t_0)}/L^{\alpha/2} \notag\\
 +\sum_{d> D}C'*e^{-\mu' d+v'(t-t_0)},
\end{align}
where $C'=C*|l_0|$, $C''=2C*|l_0|*\sqrt{918\ln2\times\abs{A}\log\abs{A}}$.
Similar to the previous arguments, even in higher dimensions, the inhomogeneous term remains negligible when the environment is sufficiently large.

In the original N-Z equation, the inhomogeneous term is difficult to consider. Hence, the approach is often limited to cases with no correlation or zero discord correlations. The N-Z equation with assignment maps  improves this limitation. As long as there is no initial entanglement, the inhomogeneous term vanishes. If the environment is sufficiently large and satisfies the conditions mentioned above, the inhomogeneous term is negligible, even if the system and environment are maximally entangled. This result is similar to diffusion. The amount of information is limited. Once the information is flowing to an infinitely  large environment,  the information rarely flows back to the system.

\section{conclusion AND OUTLOOK}\label{CON}
In this paper, we rewrite the N-Z equation with assignment maps. Compared to the projection operator method,  assignment maps provide  a smaller irrelevant part. This approach extends the application of the equation and improves  the accuracy of the expansion. We also apply the new N-Z equation to the quantum lattice model with only nearest-neighbor interactions. The special structure enables us to further analyze the impact of the inhomogeneous term. The influence of the inhomogeneous term can be described as the inflow of information from the environment. The propagation speed of  information is bounded by a finite velocity. In the short term, only a limited range of information can reach the system. Moreover, the entanglement is monogamous. Hence, if each part of the environment contains the  same amount of information, the  information flow back to the system within a finite time is negligible when the environment is sufficiently large. The negligible inhomogeneous term means that even if the system and environment are maximally entangled, the N-Z equation with assignment maps is still applicable. 

The upper bound of the inhomogeneous terms provided in this article is not very tight. It may be improved via further development of the expansion theory of operators\cite{P18}. Additionally, we consider only short-range interactions. If one can bound the inhomogeneous terms under long-range interactions, as is done in Lieb-Robinson bound, the applicability would be greatly improved. Finally, our bound  is increasing with time, but the impact of history should decrease at long time. Hence, one may find an upper bound that decreases with time. In that way, the Markov order\cite{TPMT19} can be defined directly by the N-Z equation. 
\begin{acknowledgments}
Financial support from the National Natural Science Foundation of China under Grant
Nos. 11725524, 61471356 and 11674089 is gratefully acknowledged.
\end{acknowledgments}

\end{document}